\documentclass[english,11pt,a4paper,oneside]{article}

\usepackage[T1]{fontenc}
\usepackage[utf8]{inputenc}
\usepackage{listings}
\usepackage{enumerate}
\usepackage[margin=1in]{geometry}

\usepackage{caption}
\usepackage{subcaption}

\usepackage{authblk}

\usepackage{amsthm}
\theoremstyle{definition}

\usepackage{amsfonts}
\usepackage[font=small]{caption}

\makeatletter
\usepackage{amsmath}
\usepackage{amssymb}
\usepackage{graphicx}
\usepackage{varioref}
\makeatother
\numberwithin{equation}{section}

\newcommand{\pd}{\partial}
\newcommand{\pt}{p_{t}}

\newcommand{\ppsi}{p_{\psi}}
\newcommand{\pphi}{p_{\phi}}
\newcommand{\bpsi}{b_{\psi}}
\newcommand{\bphi}{b_{\phi}}
\newcommand{\Qo}{Q_{1}}
\newcommand{\Qf}{Q_{2}}

\newcommand{\gto}{\tilde{\gamma}_1}
\newcommand{\gtt}{\tilde{\gamma}_2}
\newcommand{\mpsi}{m_{\psi}}
\newcommand{\mphi}{m_{\phi}}

\newcommand{\Lt}{\tilde{\Lambda}}

\pdfoutput=1
\usepackage{babel}
\begin{document}

\title{Geodesics in supersymmetric microstate geometries}
\author{Felicity C. Eperon}
\affil{Department of Applied Mathematics and Theoretical Physics, University of Cambridge, Wilberforce Road, Cambridge, CB3 0WA, UK \\ fce21@cam.ac.uk}

\maketitle

\begin{abstract}
It has been argued that supersymmetric microstate geometries are classically unstable. One argument for instability involves considering the motion of a massive particle near the ergosurface of such a spacetime. It is shown that the instability can be triggered by a particle that starts arbitrarily far from the ergosurface. Another argument for instability is related to the phenomenon of stable trapping of null geodesics in these geometries. Such trapping is studied in detail for the most symmetrical microstate geometries. It is found that there are several distinct types of trapped null geodesic, both prograde and retrograde. Several important differences between geodesics in microstate geometries and black hole geometries are noted. The Penrose process for energy extraction in these geometries is discussed.
\end{abstract}

\section{Introduction}
Supersymmetric microstate geometries are found as stationary solutions of type IIB supergravity that have no horizon \cite{Maldacena:2000dr,Lunin:2002iz,Giusto:2004id,Giusto:2004ip,Giusto:2004kj,Bena:2005va,Berglund:2005vb,Gibbons:2013tqa}. Several families of such solutions have been constructed, for example the 2-charge D1-D5 and 3-charge D1-D5-P supersymmetric microstate geometries of \cite{Maldacena:2000dr,Lunin:2002iz,Giusto:2004id,Giusto:2004ip,Giusto:2004kj}, which are regular except for possible orbifold singularities that can be removed in some cases. These microstate geometries can be dimensionally reduced to give solutions of 6-dimensional supergravity, although they only have 5 non-compact dimensions and are thus intended to describe microstates of 5 dimensional black holes.

A particularly important feature of these supersymmetric microstate geometries is that they admit an evanescent ergosurface \cite{Gibbons:2013tqa}, \cite{Niehoff:2016gbi}: a timelike submanifold where a causal Killing vector field $V$ is null, although it is timelike everywhere else. As solutions of 6d supergravity, the supersymmetric microstate geometries must admit a Killing vector field $V$ which is globally null. After dimensional reduction to 5d, $V$ is a causal Killing vector field; it is $V$ that defines the evanescent ergosurface for the supersymmetric microstate geometries, since it is timelike everywhere except on the evanescent ergosurface. In 5d, we can define a conserved energy along a geodesic with respect to the causal Killing vector $V$, and this is always positive. The only geodesics for which the energy is zero are null geodesics with tangent $V$ that are on the evanescent ergosurface (which is geodesic).

Recently in \cite{Eperon:2016cdd} it has been argued that the supersymmetric microstate geometries are classically unstable. The heuristic argument for instability from \cite{Eperon:2016cdd} involves a massive particle and goes as follows. Suppose we perturb the spacetime by adding a massive particle near the evanescent ergosurface that cannot escape to infinity. If we include interactions with the other fields, the particle does not follow a geodesic and radiates, losing energy. As it loses energy its trajectory tends towards the zero-energy geodesics, which are all null and on the evanescent ergosurface. Thus there is a massive particle following an almost null trajectory. To a local observer this particle will have huge energy, suggesting an instability. Thus it is interesting to consider whether we can set up the initial conditions for this process, that is, whether it is possible to have a massive particle  that exists arbitrarily far away from the evanescent ergosurface but with positive binding energy, so that it does not escape to infinity.

The second argument in \cite{Eperon:2016cdd} for non-linear instability of the supersymmetric microstate geometries is that solutions of the wave equation decay very slowly. The reason for this is that these spacetimes exhibit \textit{stable trapping}: there are null geodesics that are trapped in some bounded region of space and for which initially close geodesics remain nearby, so they are stable to small perturbations. Since the null geodesics with tangent $V$ on the evanescent ergosurface minimize the energy, they are stably trapped. Stable trapping can cause problems for the decay of solutions to the wave equation because it is possible to construct solutions that are localised near a null geodesic for an arbitrarily long time \cite{Sbierski:2013mva}.  In the most symmetrical 2- and 3-charge microstate geometries, quasinormal mode solutions were found in \cite{Eperon:2016cdd} that are localised near the stably trapped null geodesics and decay very slowly. In \cite{Eperon:2016cdd} it was shown that this leads to a particularly slow rate of decay for solutions to the wave equation in the supersymmetric microstate geometries, which was proven rigorously in \cite{Keir:2016azt}. This motivates the study of trapping in the supersymmetric microstate geometries.

The microstate geometries are supposed to be microstates of a black hole, so one might expect that they exhibit some similar properties to a black hope spacetime. It is therefore interesting to compare the geodesics of the microstate geometries to the geodesics around a Kerr black hole. The geodesics in the Kerr spacetime have been studied extensively, see for example \cite{Chandrasekhar1984}, \cite{Teo2003}. In Kerr, there are circular unstably trapped null geodesics in the equatorial plane, but no stable photon orbits: if perturbed, a photon will either fall into the black hole or escape to infinity. There are also null geodesics that are localised on spheres with radius between the radii of the unstably trapped circular geodesics in the equatorial plane. Since the microstate geometries have 5 non-compact dimensions, we should actually compare the geodesics in the microstate geometries to those around Myers-Perry black holes. These have been classified in \cite{Diemer:2014lba}; it is particularly interesting to note that although there are unstable circular geodesics, there are no \textit{stable} circular null (or timelike) geodesics in the equatorial plane. In the most symmetrical microstate geometries we will look for the equivalent of these orbits. In contrast to the black holes, we find that there are both stably and unstably trapped null geodesics (which are can be circular) in the equatorial plane as well as other constant radius null geodesics that are not necessarily in the equatorial plane. 

It has been shown in \cite{Chervonyi:2013eja} that for the most symmetrical 2-charge microstate geometries the Hamilton-Jacobi equation for null geodesics separates in certain coordinates; we find that the same happens for the most symmetrical 3-charge microstate geometries. This separability of the Hamilton-Jacobi equation is due to the fact that these spacetimes have a 'hidden' symmetry related to a conformal Killing tensor which also allows the wave equation to separate in both cases \cite{Lunin:2001dt,Giusto:2004ip}.

We will characterise the null geodesics in the most symmetrical supersymmetric microstate geometries, in particular focussing on whether there are stably trapped or unstably trapped null geodesics since these are important for decay of solutions to the wave equation. Trapping is best understood on the tangent bundle, so we will study regions of the bundle for which trapping occurs. In section \ref{sec:geo} we find that it is possible to have the massive particle for the instability mechanism discussed above for general microstate geometries. In section \ref{sec:2charge} we consider the 2-charge case: after separating the Hamilton-Jacobi equation we will investigate the null geodesics with zero momentum around the internal torus (these give also correspond to null geodesics in 6 dimensions) in both of the submanifolds $\theta=\pi/2$ and $r=0$ and show that there are both stable and unstable trapped geodesics. We will also find geodesics that remain at constant radius, and are thus trapped, but not necessarily in the equatorial plane. In addition, in section \ref{sec:timelike} we consider the geodesics with non-zero momentum around the internal torus (these give massive particles in 5 or 6 dimensions) in the equatorial plane to support the heuristic argument for instability in \cite{Eperon:2016cdd} that involves a massive particle. The  3-charge case is more complicated, but in section \ref{sec:3charge} we find that the Hamilton-Jacobi equation separates and that in general there are both stable and unstable trapped geodesics, as well as constructing an example of the Penrose process to extract energy in these spacetimes.

\section{Geodesics of general microstate geometries} \label{sec:geo}
For a general supersymmetric microstate geometry, we start off in 10 dimensions before reducing to 6 dimensions by compactifying on the internal torus. Suppose we have a high energy graviton in 10d with non-zero momentum around the internal torus. Since it is high energy, we can use the geometric optics approximation to say that we expect it to be localised near a null geodesic and therefore investigate the null geodesics in 10d. If the geodesics have non-zero momentum on the internal torus they correspond to trajectories of massive particles after dimensional reduction to 6d. In the Introduction we described the mechanism for instability from \cite{Eperon:2016cdd} that involved a massive particle with a bound  trajectory\footnote{\textit{Bound} means that the particle cannot escape to infinity; since there is no black hole for it to fall in to this is the same as the particle following a geodesic that is \textit{trapped}. However, when we talk about massive particles in 6d we will use 'bound', and use 'trapped' as a description of the null geodesics.}. We will establish whether it is possible in a general supersymmetric microstate geometry to find such a particle with positive binding energy arbitrarily far away that does not escape to infinity. Note that this is not necessarily obvious, although gravity is an attractive force, because there are various conformal factors involved in the dimensional reduction that have an effect on the particle. To find the geodesics in the spacetime we will use the Hamilton-Jacobi equation.

The Hamilton-Jacobi function $S$ for geodesics satisfies \begin{equation} g^{a b}\nabla_{a}S\nabla_{b}S=-\mu^2.\label{eq:HJ} \end{equation} where $\mu=0$ for null geodesics or $\mu$ is the mass of the particle following a timelike geodesic. This implies that \begin{equation} P_{a}=\nabla_{a}S \label{eq:u} \end{equation} is the momentum of a particle following a causal geodesic: \begin{equation} P^{a}\nabla_{a}P_{b}=P^{a}\nabla_{a}\nabla_{b}S=P^{a}\nabla_{b}\nabla_{a}S=P^{a}\nabla_{b}P_{a}=\frac{1}{2}\nabla_{b}(P^{a}P_{a})=0 \end{equation} and $g^{ab}P_aP_b=-\mu^2$.

\subsection{10d null geodesics}\label{sec:generalHJ}
We will look for null geodesics in the full 10 dimensions of solutions to type IIB supergravity compactified on $T^4$. We can write the string frame metric as \begin{equation}
(g_{10}^S)_{\hat{\mu}\hat{\nu}}dx^{\hat{\mu}}dx^{\hat{\nu}}=(g_{6})_{\mu \nu}dx^{\mu}dx^{\nu}+e^{2\Psi}\delta_{ij}dy^idy^j \label{eq:generalmetric}
\end{equation} where $\mu,\,\nu=0\dots 5$, $\hat{\mu},\,\hat{\nu}=0\dots 9$ and $i,\,j=1\dots 4$, $y^i$ are the coordinates on the internal torus and $\Psi(x)$ is some function independent of these coordinates. The upper index $S$ refers to the fact that this metric is in the string frame. To obtain the 6d Einstein frame metric, one dimensionally reduces on $T^4$, which gives rise to one conformal factor, and then multiplies by a conformal factor to go from the string to Einstein frame; when we go from 10 to 6 dimensions it turns out that these conformal factors cancel. This means $g_6$, the 6d part of the 10d string frame metric in \eqref{eq:generalmetric} is in fact the 6d Einstein frame metric.

The Hamilton-Jacobi equation for null geodesics in 10d is \begin{equation} 
0=g_{10}^{\hat{\mu}\hat{\nu}}\pd_{\hat{\mu}} S\pd_{\hat{\nu}}S=g_{6}^{\mu \nu}\pd_{\mu} S\pd_{\nu}S+e^{-2\Psi}\delta^{ij}\pd_{i} S\pd_{j}S. \label{eq:HJsplit} \end{equation}

Let \begin{equation}
S=\tilde{S}(x^{\mu})+q_iy^i 
\label{eq:S10}
\end{equation}	
where $y^i$ are coordinates on the internal torus, $x^{\mu}$ are the other coordinates and $q_i=\left(\pd/\pd y^{i}\right)\cdot P$ are the conserved quantities associated with the Killing vectors $\pd/\pd y^{i}$ along a geodesic with momentum $P_a$.

Substituting \eqref{eq:S10} in \eqref{eq:HJsplit} gives \begin{equation}
g_{6}^{\mu \nu}\pd_{\mu} \tilde{S}\pd_{\nu}\tilde{S}=-e^{-2\Psi}\mu^2, \qquad \text{ where } \mu^2=\delta^{ij}q_i q_j \label{eq:HJ6einstein} 
\end{equation} which implies that $\tilde{S}$ satisfies the Hamilton-Jacobi equation with metric $\tilde{g}_6$: \begin{equation}
\tilde{g}_{6}^{\mu \nu}\pd_{\mu} \tilde{S}\pd_{\nu}\tilde{S}=-\mu^2, \label{eq:HJ6} \end{equation} where the rescaled 6d metric is \begin{equation}
(\tilde{g}_6)_{\mu \nu}=e^{-2\Psi}({g}_6)_{\mu \nu}. \label{eq:gtilde} \end{equation}

If all the momenta around the internal torus are zero so that $\mu=0$, this is exactly the Hamilton-Jacobi equation for null geodesics in 6d since the conformal factor in front of the Einstein frame metric has no effect. If $q_i\neq 0$, this is the Hamilton-Jacobi equation for timelike geodesics with mass $\mu$, \textit{not} for the 6d Einstein frame metric but for $\tilde{g}_6$, the Einstein frame metric multiplied by a conformal factor. 

Now suppose we have a general isolated gravitating system in (5+1)-dimensions where one of the dimensions is compact and near infinity it looks like a product of (4+1)-d Minkowski space with a Kaluza-Klein $S^1$. We verify that for these spacetimes there are trajectories with $\mu\neq 0$ that start arbitrarily far from the evanescent ergosurface but cannot escape to infinity. This includes general supersymmetric microstate geometries, so we can then apply the argument for instability from the Introduction. In the full 10 dimensions, these geodesics correspond to null geodesics with non-zero momentum around the internal torus that are stably trapped.

We will assume that the spacetime is stationary and that $T=\pd/\pd t$ is the Killing vector field that is timelike near infinity. Then the energy $E=-T\cdot P=-P_t$ is conserved along geodesics. If the spacetime is a solution of supergravity there is also a globally null Killing vector field $V$. In the supersymmetric microstate geometries, $V=T+Z$ where $z$ is the coordinate around the Kaluza-Klein $S^1$ and  $Z=\pd/\pd z$ is a Killing vector field. We can define a positive conserved energy with respect to $V$ both in 10d $E_{10}=-V\cdot P$, and in 6d, $E_6=-V^aP_a=-P_t-P_z$. Note that in 6d $P$ is tangent to a causal geodesic of $\tilde{g}_6$ but we still have $E_6=E_{10}$, which is clear when we write $E_{10}=-V^aP_a$, as the conformal factors for the metrics in different frames all cancel out and the energy is conserved from both the 10d and 6d perspectives. Note that in 6d the $T$ energy is positive, $E\geq0$, when $P_z=0$.

To proceed, we show that under certain conditions it is always possible to find timelike vectors of the conformal metric $\tilde{g}_6=e^{-2\Psi}g_6$ ($g_6$ is the Einstein frame metric) arbitrarily far out, that satisfy $\tilde{g_6}^{\mu\nu}P_{\mu}P_{\nu}=-\mu^2$ with $E<\mu$. If we can find such a vector at a point $p$ in the spacetime then we can find a timelike geodesic (of $\tilde{g}$) through $p$ with momentum $P$ which is bound, since the binding energy $\mu-E$ is positive. 

The asymptotic form of the Einstein frame metric for an isolated gravitating system in (4+1) dimensions is given in \cite{Myers:1986un}. We can extend this to our case by including an extra compact dimension. To calculate the asymptotic form of the 6d metric we have to dimensionally reduce to 5d and then compare these metric components to those of the general asymptotic form in \cite{Myers:1986un}. 

To reduce from the 6 to 5d Einstein frame metric, \begin{equation}
g_6=e^{2a_1\Phi}g_5+e^{2a_2\Phi}(dz+\mathcal{A})^2 \label{eq:6to5}
\end{equation} where $a_1^2=1/40$, $a_2=-4a_1$, $z$ is the coordinate around the extra compact dimension and $\Phi(x)$ is a scalar field. We assume that \begin{equation}
\Phi=\frac{b}{r^2}+O(r^{-3}),\qquad \mathcal{A}_{\alpha}=\frac{c_{\alpha}}{r^2}+O\left(\frac{1}{r^3}\right)
\end{equation}  as $r\rightarrow \infty$ for some constants $b,\,c_{\alpha}$, ($\alpha=0\dots4$), since the spacetime is asymptotically  $\mathcal{M}_{4,1}\times S^1$.

In 5d, we have a general isolated gravitating system and can use the results of \cite{Myers:1986un}. The system is asymptotically flat, so near infinity we write \begin{equation}
(g_5)_{\alpha\beta}=\eta_{\alpha\beta}+\hat{h}_{\alpha\beta}
\end{equation} where $|\hat{h}_{\alpha \beta}|\ll 1$ and $\alpha,\beta\in\{t,i\}$, $i=1\dots 4$. We have that \begin{equation}
\hat{h}_{tt}=\frac{d_1}{r^2},\qquad \hat{h}_{ij}=\frac{d_2}{r^2}\delta_{ij},\qquad \hat{h}_{ti}=O\left(\frac{1}{r^3}\right)
\end{equation} for some positive constants $d_1,\,d_2$ that can be found in \cite{Myers:1986un}.

Now in 6d, we can write the metric near infinity as \begin{equation}
(g_6)_{\mu \nu}=\eta_{\mu\nu}+h_{\mu\nu}
\end{equation} where, from \eqref{eq:6to5}, at leading order, \begin{equation}
h_{tt}=\frac{e_1}{r^2},\qquad h_{ij}=\frac{e_2}{r^2}\delta_{ij},\qquad h_{zz}=\frac{e_3}{r^2},\qquad h_{\alpha z}=\frac{c_{\alpha}}{r^2},\qquad h_{ti}=O\left(\frac{1}{r^3}\right) \label{eq:h}
\end{equation} where $e_1=d_1-2a_2 b$, $e_2=d_2+2a_2b$, $e_3=2a_2 b$, $i,j=1\dots 4$ and $\alpha \in\{t,i\}$.

The inverse metric at this order is $g^{\mu\nu}=\eta^{\mu\nu}-h^{\mu\nu}$ where the indices on the right hand side are raised with $\eta^{\mu\nu}$.

We assume that for large $r$, \begin{equation}
e^{2\Psi}=1+\frac{f}{r^2}+O\left(\frac{1}{r^3}\right) \label{eq:conformal}
\end{equation} for some constant $f$. 

Timelike vectors of $\tilde{g}$ must satisfy \begin{equation}
\tilde{g}^{\mu\nu} P_{\mu}P_{\nu}=e^{2\Psi}g^{\mu\nu}P_{\mu}P_{\nu}=-\mu^2. \label{eq:timelikevector}
\end{equation}

Substituting \eqref{eq:h} and \eqref{eq:conformal} into \eqref{eq:timelikevector} gives \begin{equation}
(-1-\frac{f+e_1}{r^2})P_t^2+(1+\frac{f-e_2}{r^2})\sum_{i=1}^{4}P_i^2+(1+\frac{f-e_3}{r^2})P_z^2+2(\frac{c_t}{r^2}P_t-\sum_{i=1}^{4}\frac{c_i}{r^2}P_i)P_z +O(r^{-3})=-\mu^2. \label{eq:timelike2}
\end{equation}

If $P_z=0$, solving for $P_t$ gives \begin{equation}
P_t^2=\mu^2-\frac{f+e_1}{r^2}\mu^2-(1-\frac{e_1+e_2}{r^2})\sum_{i=1}^4P_i^2+O(r^{-3}). \label{eq:Pt2}
\end{equation}
Therefore, if \begin{equation} f+e_1>0 \end{equation} it is possible to choose $P_i$ small enough such that \eqref{eq:Pt2} is satisfied and that $P_t^2<\mu^2$.

The condition that the particle has positive binding energy, $E^2<\mu^2$, implies that the particle cannot escape to infinity and is stably trapped. To see this, take $r\rightarrow\infty$ in \eqref{eq:timelike2}: the left hand side  $LHS\geq -E^2$ while the right hand side $RHS<-E^2$, since these cannot be equal it is not possible to have timelike vectors with $E^2<\mu^2$ as $r\rightarrow\infty$. Along a geodesic, $E$ is conserved, so the geodesic can never go all the way out to infinity. Furthermore, even if the geodesic is perturbed slightly the binding energy is positive so it still will not escape to infinity. Thus the geodesic is stably trapped.

For the general 2-charge microstate geometries in \cite{Lunin:2002iz}, \begin{equation}
f+e_1=\Qo>0
\end{equation} provided $\Qo>0$, which is indeed the case for the microstate geometries in \cite{Lunin:2002iz}. For the supersymmetric 3-charge solutions in \cite{Giusto:2004ip},  $f+e_1=\Qo+Q_p>0,$ since $Q_1>0$ and $Q_p>0$ in these solutions. This implies that it is always possible to find a stably trapped null geodesic with non-zero momenta around the internal torus in these geometries. Moreover, this is possible arbitrarily far out and thus we can set up the initial conditions for the heuristic argument of instability given in the Introduction by the presence of stably trapped massive particles. If the asymptotics only depend on the charges, we expect this to also be the case for all of the 2-charge solutions.

\section{2-charge microstate geometry}\label{sec:2charge}
In this section we study the trapping of \textit{null} geodesics in the most symmetrical supersymmetric microstate geometries. These geodesics are of interest for the second argument for instability in the Introduction, involving slow decay of solutions of the wave equation. 

We consider the smooth 2-charge supersymmetric microstate geometries constructed in \cite{Maldacena:2000dr,Lunin:2002iz}. They are supersymmetric solutions of type IIB supergravity compactified on $T^4$ with two charges arising from $n_1$ D1-branes wrapped around a Kaluza-Klein $S^1$ and $n_2$ D5-branes wrapped around $S^1\times T^4$. It is possible to dimensionally reduce both the 2-charge and 3-charge microstate geometries to 5 dimensions but we will study them in 6d because they are regular in 6d but not in 5d. 

The 10d string frame metric for this 2-charge D1-D5 microstate geometry (in the form given in \cite{Lunin:2001dt}) is \begin{equation}\begin{split} ds^2_{10}=& -\frac{1}{h}(dt^2-dz^2)+hf\left(d\theta^2+\frac{dr^2}{r^2+a^2}\right)-\frac{2a \sqrt{\Qo\Qf}}{hf}\left(\cos^2\theta dz d\psi+\sin^2\theta dt d\phi\right) \\& + h\Big[\left(r^2+\frac{a^2\Qo\Qf\cos^2\theta}{h^2f^2}\right)\cos^2\theta d\psi^2+\left(r^2+a^2-\frac{a^2\Qo\Qf\sin^2\theta}{h^2f^2}\right)\sin^2\theta d\phi^2\Big]\\& +\sqrt{\frac{H_1}{H_2}}\Sigma_{i=1}^{4}dx_i^2
\\=& ds^2_6+\sqrt{\frac{H_1}{H_2}}\Sigma_{i=1}^{4}dx_i^2 \end{split} \label{eq:2metric} \end{equation}
where $r\geq0,\,\theta\in[0,\pi/2]$, $0 \leq \phi,\psi\leq 2\pi$, $z\sim z+2\pi R_z$, \begin{equation}
f=r^2+a^2\cos^2\theta,\qquad H_i=1+\frac{Q_i}{f}\;\;\;(i=1,\,2),\qquad h=\left(H_1 H_2\right)^{\frac{1}{2}}
\end{equation} and the solution is written in terms of the charges \begin{equation}
\Qo=\frac{(2\pi)^4g\alpha'^3}{V}n_1 \qquad \Qf=g\alpha'n_2 
\end{equation}
where $g$ is the string coupling constant, $V$ is the volume of the $T^4$ and the length scale $a$ is defined by
\begin{equation}
a =\frac{\sqrt{Q_1Q_2}}{R_z}. \label{eq:a}
\end{equation}
There is a globally null Killing vector field \begin{equation}
V=\frac{\pd}{\pd t}+\frac{\pd}{\pd z}.
\end{equation}

As in \cite{Eperon:2016cdd}, we define the evanescent ergosurface $\mathcal{S}$ in 6d to be the locus of points where $V\cdot Z=0$, where $Z=\pd/ \pd z$ is the Kaluza-Klein Killing vector field; in 5d $V$ is timelike everywhere except on the evanescent ergosurface, where it is null. Therefore the evanescent ergosurface is given by $f=0$, i.e. $r=0$ and $\theta=\pi/2$, with topology $S^1$ at constant $t$.

This is a particularly important submanifold since it was shown in \cite{Eperon:2016cdd} that there are stably trapped null geodesics with zero energy that stay on the evanescent ergosurface. Indeed, the only geodesics with zero Kaluza-Klein momentum that have zero energy are those with tangent $V$ that are stably trapped on $\mathcal{S}$. It was also shown in \cite{Eperon:2016cdd} that geodesics with tangent $V$ through any point in the spacetime are stably trapped, but those away from $\mathcal{S}$ have non-zero Kaluza-Klein momentum.

\subsection{Hamilton-Jacobi equation}
The Hamilton-Jacobi equation for null geodesics (in 10d) is separable due to a conformal Killing tensor \cite{Chervonyi:2013eja}. We can then separate the equation for geodesics in 6d, so in \eqref{eq:HJ6} let
\begin{equation}
\tilde{S}=p_{I}x^{I}+R(r)+\Theta(\theta). 
\label{eq:S6} 
\end{equation} where $p_I=\left(\pd/\pd x^{I}\right)\cdot P$.		
		
Using the ansatz \eqref{eq:S6} equation \eqref{eq:HJ6} separates into equations for $R(r)$ and $\Theta(\theta)$. Then recall that $P_{\mu}=\nabla_{\mu}S$ is the momentum of a particle following a null geodesic so $P^{\mu}=\dot{x}^{\mu}(\lambda)=g_{10}^{\mu\nu}\nabla_{\nu}S$ where $\lambda$ is an affine parameter along the geodesic, since these geodesics are null in 10d and we can rescale the affine parameter along the null geodesic. Then using  \begin{equation*}
\dot{r}(\lambda)=\frac{dr}{d\lambda}=g_{10}^{r\mu}\pd_{\mu} S=g_{10}^{rr}\pd_r R,\qquad \dot{\theta}(\lambda)=\dfrac{d\theta}{d \lambda}=g_{10}^{\theta \mu}\pd_{\mu} S=g_{10}^{\theta \theta}\pd_{\theta}\Theta
\end{equation*} where $\lambda$ is an affine parameter along the geodesic, we find coupled first order differential equations for $\dot{\theta}$ and $\dot{r}$:

\begin{align}
\dot{r}^2+\frac{1}{(hf)^2}V_r(r)=0 \label{eq:req} \\
\dot{\theta}^2+\frac{1}{(
	hf)^2}V_{\theta}(\theta)=0 \label{eq:theq}
\end{align}
where the effective potentials are \begin{equation}\begin{split}
V_r=(p_z^2-p_t^2+&\mu^2)r^4+ \left(a^2(p_z^2-\pt^2+\mu^2)+\Lambda+ \mu^2\Qf \right)r^2-(a\pphi+\sqrt{\Qo\Qf}p_t)^2\\&+(a\ppsi+\sqrt{\Qo\Qf}p_z)^2 +\Lambda a^2 + \mu^2a^2\Qf +(a\ppsi+\sqrt{\Qo\Qf}p_z)^2\frac{a^2}{r^2} \end{split} \label{eq:Vr} \end{equation}
and 
\begin{equation}
V_{\theta}=\frac{\pphi^2}{\sin^2\theta}+a^2\cos^2\theta(p_z^2-\pt^2+\mu^2)+\frac{\ppsi^2}{\cos^2 \theta}+ (\Qo+\Qf)(p_z^2-\pt^2)-\Lambda \label{eq:Vth}
\end{equation} where $\Lambda$ is a constant arising from the separation of variables.
	
Equations \eqref{eq:req} and \eqref{eq:theq} are coupled via the factor $hf(r,\theta)$ so that they are not in fact in the form of 1d equations of motion with an effective potential. However, this factor is strictly positive so we can still say something about the trapped geodesics by investigating the signs of $V_r$ and $V_{\theta}$ since geodesics can only exist in regions where $V_r(r)\leq 0$ and $V_{\theta}(\theta)\leq0$.

\subsubsection{Trapping}\label{sec:trapping}

For a geodesic to exist there must be some value of $\theta$ such that $V_{\theta}(\theta)\leq 0$. Assuming that this is the case, whether or not there is trapping depends only on the radial effective potential $V_r$. 

\begin{figure}[h]
	\centering
	\begin{minipage}{0.49\textwidth}
		\centering	
		\includegraphics[width=0.9\textwidth]{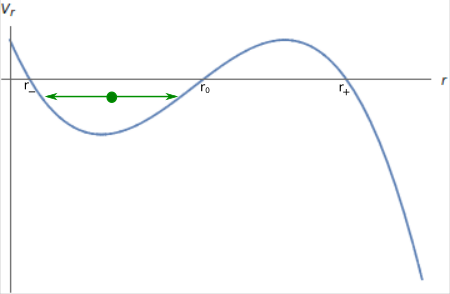} 
		\caption{Stable trapping}
		\label{fig:trappingstable}
	\end{minipage}\hfill
	\begin{minipage}{0.49\textwidth}
		\centering
		\includegraphics[width=0.9\textwidth]{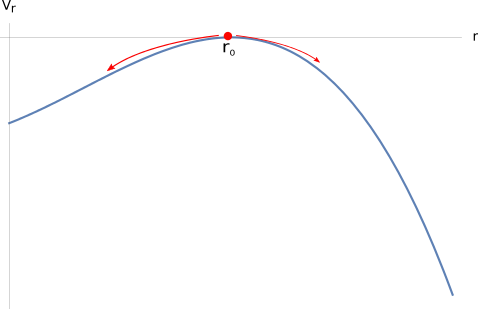}
		\caption{Unstable trapping}
		\label{fig:trappingunstable}
	\end{minipage}
\end{figure}

Stable trapping occurs if there are some values $r_-\leq r_0< r_+$ such that $V_r(r)\leq 0$ for $r_-\leq r\leq r_0$ but $V_r(r)>0$ for $r_0<r<r_+$; see Figure \ref{fig:trappingstable} for an example. The geodesics are then allowed to propagate in $r_-\leq r\leq r_0$ but cannot escape to infinity due to the potential barrier at $r_0<r<r_+$. Suppose there is a geodesic that is trapped in this region. If we perturb this trapped geodesic (in the tangent bundle) the shape of the potential changes slightly but there will still be a potential barrier and the perturbed geodesic is also trapped. This is stable trapping since perturbing a trapped geodesic can only give another trapped geodesic.  

On the other hand, there is unstable trapping if there is some value $r_0$ such that $V_r(r)$ has a (local) maximum at $r_0$ and $V_r(r_0)=0$, as in Figure \ref{fig:trappingunstable}: the geodesic at $r_0$ stays there, but if there is a nearby geodesic at $r=r_0+\epsilon$ for some $\epsilon$ then that geodesic can escape away from $r_0$.  Also, if we perturb the potential slightly there will no longer be a maximum when $V_r=0$  and there is no reason for geodesics to stay near there: if the maximum moves in such a way to be at a negative value of $V_r$ the perturbed geodesics are not trapped but can escape out to infinity, so this is unstable trapping. Note that the unstably trapped geodesic will always remain at constant $r$, so that in the equatorial plane $\theta=\pi/2$ these orbits are all circular. 

We can distinguish between trapped null geodesics with positive or negative angular momentum (in the $\phi-$direction), $\pphi>0$ or $\pphi<0$, in the same way as for the unstably trapped circular null geodesics in the equatorial plane of Kerr black holes, see for example \cite{Bardeen:1972, Chandrasekhar1984}. In Kerr (taking $a=J/M>0$), there are two possible radii for circular orbits in the equatorial plane: at $r=r_-$ there are \textit{direct} or \textit{prograde} circular orbits with positive angular momentum $\pphi>0$ while at $r=r_+$ there are \textit{retrograde} circular orbits with negative angular momentum $\pphi<0$. Note that this definition of pro- and retrograde orbits applies when the angular momentum of the spacetime is positive ($a>0$). Since we also have positive angular momentum (in the $\phi-$direction) $J_\phi>0$, we will define prograde and retrograde orbits in the same way. 

For geodesics tangent to $V$, we can calculate $\pphi=-2a\sqrt{\Qo\Qf}/hf<0$ and see that all of these geodesics, including those with zero energy on the evanescent ergosurface, are retrograde. This is best explained in 5d: as $V$ approaches a standard time translation at infinity the geodesics with tangent $V$ do not rotate with respect to infinity. However, $J_{\phi}>0$ so the spacetime has non-zero, positive angular momentum. This means that these geodesics are resisting the frame-dragging effect of the geometry and thus have angular momentum opposite to that of the spacetime.

\subsubsection{Null geodesics and solutions of the wave equation}
In the geometric optics approximation, one can find solutions of the wave equation $\Box \Phi =0$ that are localised near null geodesics. Indeed, this was made precise in \cite{Sbierski:2013mva}, where it was shown that it is possible to construct solutions of the wave equation that are localised near a null geodesic for an arbitrarily long time, and for which the energy of the solution is close to the conserved energy of the null geodesic. This implies that trapping can lead to slow decay of solutions to the wave equation.

We can use certain properties of the null geodesics to say something about the corresponding solutions of the wave equation. For instance, in Kerr there are quasinormal modes localised near the unstably trapped circular photon orbits in the equatorial plane. In ref. \cite{Yang:2012he} it is shown that not only can we approximate the energy of the solutions by the energy of the null geodesic, but we can also approximate the rate of decay of the mode using the rate at which neighbouring geodesics spread out away from the circular one. In the supersymmetric microstate geometries we have stable trapping and therefore we expect that the corresponding quasinormal modes decay far slower than in Kerr because the nearby geodesics remain close to the stably trapped one; this was indeed shown to be the case and to lead to very slow decay of solutions to the wave equation in \cite{Eperon:2016cdd}, which was proved rigourously in \cite{Keir:2016azt}.

\subsection{6d null geodesics in equatorial plane $\theta=\pi/2$}\label{sec:pi/2}
We expect stable trapping near (in the tangent bundle) the zero energy null geodesics that are stably trapped on the evanescent ergosurface and we determine the region filled by such geodesics, which are retrograde. We also find that for certain microstate geometries, it is possible to have stably trapped prograde null geodesics.

We consider the null geodesics of the 6d metric that stay in the equatorial plane. Null geodesics in 10d with zero momentum around the internal torus are also null geodesics in 6d and have $\mu=0$ in equations \eqref{eq:HJ6}, \eqref{eq:Vr} and \eqref{eq:Vth}.

Although equations \eqref{eq:req} and \eqref{eq:theq} are coupled we can still find geodesics that stay at $\theta=\pi/2$, which we expect by symmetry. Differentiating equation \eqref{eq:theq} wrt $\lambda$ and dividing through by $\dot{\theta}$ gives the second order equation for $\theta(\lambda)$: \begin{equation}
2\ddot{\theta}-hf\dot{r}\sqrt{-V_{\theta}}\pd_r\left((hf)^{-2}\right)+V_{\theta}\,\pd_{\theta}\left((hf)^{-2}\right)+(hf)^{-2}\pd_{\theta}V_{\theta}=0. \label{eq:thetaddot}
\end{equation}	
Suppose that for some $\lambda_0$ we have $\theta(\lambda_0)=\theta_0$ and $\dot{\theta}(\lambda_0)=0$ so that $V_{\theta}(\theta_0)=0$; if also $V_{\theta}'\left(\theta_0\right)=0$ equation \eqref{eq:thetaddot} implies that $\ddot{\theta}(\lambda_0)=0$ and so the geodesic remains at constant $\theta_0$. 

For geodesics at $\theta=\pi/2$ we must have $V_{\theta}(\pi/2)=0$, i.e. \begin{equation}
\ppsi=0, \hspace{5mm} \Lambda=\pphi^2+(\Qo+\Qf)(p_z^2-p_t^2). \label{eq:2val}
\end{equation} 	
Differentiating \eqref{eq:Vth} wrt $\theta$ and substituting in $\ppsi=0$, we find that we do indeed have $V_{\theta}'(\pi/2)=0$ as required for the geodesics to stay at $\theta=\pi/2 $. 
	
\subsubsection{Radial equation}
We will find the geodesics in the submanifold $\theta=\pi/2$ with zero Kaluza-Klein momentum $p_z=0$, since these geodesics will also correspond to massless particles after dimensional reduction to 5d. 

Define the impact parameters \begin{equation}
b_{\phi}=-\frac{\pphi}{p_t}, \qquad b_{\psi}=-\frac{\ppsi}{p_t},\qquad b_{z}=-\frac{p_z}{p_t}.
\end{equation} Due to the freedom to rescale the affine parameter along the geodesic, it is only these ratios that have any physical importance. We will look for values of $b_{\phi}$ that give either stable or unstable trapping ($\bpsi=0$ for these geodesics with $\theta=\pi/2$, and by our choice $b_z=0$). 

As $\pd/\pd t$ is everywhere causal in the 2-charge microstate geometry, $p_t\leq 0$ for a future-directed null geodesic and an equivalent definition of pro/retro-grade in terms of $\bphi=-\pphi/p_t$ is that an orbit is direct or prograde if $\bphi>0$ and retrograde if $b_{\phi}<0$. The zero energy null geodesics that are stably trapped on $\mathcal{S}$ have $\bphi=-\infty$.

Substituting \eqref{eq:2val} into \eqref{eq:Vr} we can write the potential as \begin{equation} \begin{split}
V_r=p_t^2(-r^4+& Br^2+C) \\ \text{ where } B=\bphi^2-(\Qo+\Qf)-a^2, \hspace{5mm} & C=2a\sqrt{\Qo\Qf}(\bphi-\xi) \label{eq:Vrpi2} \end{split} \end{equation} where \begin{equation} 
\xi=\left(\Qo\Qf+a^2(\Qo+\Qf)\right)/(2a\sqrt{\Qo\Qf})  \label{eq:xi}. \end{equation}

Note that $p_t=0$ gives the stably trapped zero energy null geodesics on the evanescent ergosurface. As discussed in section \ref{sec:trapping}, trapping depends on the sign of $V_r$. To find regions where $V_r$ is negative we note that $V_r\rightarrow -\infty$ as $r\rightarrow \infty$ so there is always an allowed region near infinity (unless $p_t=0$) and then we only have to find the roots of $V_r$, which is a quadratic polynomial in $r^2$. 

We find various different possibilities for the types of geodesic according to the values of $B$ and $C$, which depend on $\bphi$. The bounds and equations for $B$ and $C$ which lead to the different types of trapping give conditions on $\bphi$.

Define \begin{equation}
b_{\phi}^-=-a-\sqrt{\Qo}-\sqrt{\Qf} \label{eq:bphi-},
\end{equation} 
\begin{equation}
b_{\phi}^+=\max\{-a+\sqrt{\Qo}+\sqrt{\Qf},\,a\pm(\sqrt{\Qo}-\sqrt{\Qf})\},
\end{equation} 
\begin{equation} \begin{split}
&r_-^2=\max\{0,\,-\sqrt{\Qo\Qf}+ a|\sqrt{\Qo} -\sqrt{\Qf}|,\,\sqrt{\Qo\Qf}-a(\sqrt{\Qo}+\sqrt{\Qf})\}\\ &r_+^2=\sqrt{\Qo\Qf}+ a(\sqrt{\Qo}+\sqrt{\Qf}). \label{eq:radii}
\end{split} \end{equation} and
\begin{equation}
2r_1^2=\bphi^2-\Qo-\Qf-a^2-\left( (\bphi^2-\Qo-\Qf-a^2)^2+8 a \sqrt{\Qo\Qf}(\bphi-\xi)\right)^{1/2}. \label{eq:r1}
\end{equation}

There are always stably trapped retrograde geodesics with $\bphi<\bphi^-$, but what happens for larger $\bphi$ depends on the background parameters. The two different cases, which depend on the value of $R_z$ compared to $Q_i$, are:\begin{enumerate}[1.]
	\item{$(\Qo+\Qf+a^2)^{\frac{1}{2}}<\xi$}: \\ This happens if $R_z>\sqrt{\Qo}+\sqrt{\Qf}$ or $0<R_z<|\sqrt{\Qo}-\sqrt{\Qf}|$.
	\item{$(\Qo+\Qf+a^2)^{\frac{1}{2}}\geq\xi$}:\\
	This requires $|\sqrt{\Qo}-\sqrt{\Qf}|\leq R_z\leq \sqrt{\Qo}+\sqrt{\Qf}$.
\end{enumerate} 

We summarize the possible geodesics in the two cases by giving the ranges of $\bphi$ that give rise to different types of trapping:
\begin{enumerate}
	\item{$(\Qo+\Qf+a^2)^{\frac{1}{2}}<\xi$}
	\begin{enumerate}
		\item{$\bphi<\bphi^-$}: retrograde stably trapped geodesics in a region $0\leq r\leq r_1$ that includes the evanescent ergosurface at $r=0$ and $r_1$ is given in \eqref{eq:r1}.
		\item{$\bphi=\bphi^-$}: unstable trapping of retrograde geodesics at $r_+$.
		\item{$\bphi^-<\bphi<\bphi^+$}: no trapping, and geodesics from infinity can reach the evanescent ergosurface at $r=0$.
		\item{$\bphi=\bphi^+$}: prograde unstably trapped geodesics at $r_-$.
		\item{$\bphi^+<\bphi\leq \xi$}: stable trapping in the region $0\leq r<r_-$ that includes the evanescent ergosurface, and the geodesics are prograde. If $\bphi= \xi$ the geodesics are stably trapped precisely on the evanescent ergosurface at $r=0$.
		\item{$\bphi>\xi$}: no trapping, and the  geodesics are bounded away from the evanescent ergosurface.
	\end{enumerate}	
	\item{$(\Qo+\Qf+a^2)^{\frac{1}{2}}\geq \xi$}
	\begin{enumerate}
		\item{$\bphi<\bphi^-$}: retrograde stably trapped geodesics in a region $0 \leq r\leq r_1$ that includes the evanescent ergosurface.
		\item{$\bphi=\bphi^-$}: unstable trapping of retrograde geodesics at $r_+>0$.
		\item{$\bphi^-<\bphi<\xi$}: no trapping, and geodesics can reach the evanescent ergosurface.
		\item{$\bphi=\xi$}: unstable trapping of prograde geodesics on the evanescent ergosurface at $r=0$.
		\item{$\bphi>\xi$}: no trapping, but geodesics are bounded away from the evanescent ergosurface.
	\end{enumerate}
\end{enumerate}

\begin{figure}[h!]
	\centering
	\begin{minipage}{0.49\textwidth}
		\includegraphics[width=0.99\linewidth]{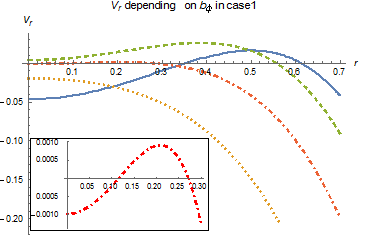}
		\caption{$V_r$ when $(\Qo+\Qf+a^2)^{\frac{1}{2}} <\xi$ with $a=1/8,\,\Qo=\Qf=1/8$. In order of increasing $\bphi$, the blue solid line is case (a), the orange dotted line is (c), the red dot-dashed line is (e) and the green dashed line is (f). The inset shows (e) zoomed in near the origin. }
		\label{fig:Vr1}
	\end{minipage}
	\hfill
	\begin{minipage}{0.48\textwidth}
		\centering
		\includegraphics[width=0.98\linewidth]{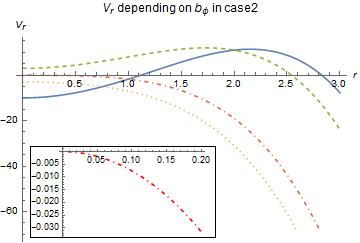}
		\caption{$V_r$ when $(\Qo+\Qf+a^2)^{\frac{1}{2}} \geq \xi $ with $a=1,\,\Qo=\Qf=1$. In order of increasing $\bphi$, the blue solid line is case (a), the orange dotted line is (c), the red dot-dashed line is (d) and the green dashed line is (e). The inset shows (d) zoomed in near the origin.}
		\label{fig:Vr2}
	\end{minipage}
\end{figure}

The effective potentials for the possible types of trapping are illustrated in Figures \ref{fig:Vr1} and \ref{fig:Vr2}.

The stably trapped geodesics are in the region $0\leq r<r_1$ where $r_1$ is given in \eqref{eq:r1}. For the prograde stably trapped geodesics this decreases as $\bphi$ increases and so the maximum radius for these geodesics is $r_-$, the radius of the unstable trapped orbits. For the retrograde stably trapped orbits, we have $\frac{dr_1^2}{d\bphi}\rightarrow \infty $ as $\bphi\rightarrow b_{\phi *}$ and $\frac{dr_1^2}{d\bphi} \rightarrow 0_+$ as $\bphi\rightarrow -\infty$; generically we expect that $\frac{dr_1^2}{d\bphi}>0$ for $-\infty <\bphi \leq b_{\phi_*}$ and thus $r_1$ decreases as $\bphi $ decreases and the maximum radius of these orbits is $r_+$, the radius of the unstably trapped retrograde orbit.

We have so far taken $\bphi$ to be finite. However, the stably trapped geodesics on the evanescent ergosurface in \cite{Eperon:2016cdd} have zero energy. If we take the limit $p_t\rightarrow 0$ in \eqref{eq:Vrpi2} the potential has the form \begin{equation}
	V_r=\pphi^2 r^2-2 a \pphi p_t +\dots 
\end{equation} where the dots represent terms for which the coefficients are $O(p_t^2)$. If $\pphi<0$ so that we are taking $\bphi\rightarrow-\infty$ the geodesics are trapped in a region near $r=0$ that becomes smaller in the limit until we have the retrograde zero energy geodesics that are stably trapped exactly at the evanescent ergosurface (this is included in case 1(a) or 2(a)). If $\pphi>0$ there could only be geodesics when $p_t=0$ which implies that these are tangent to $V$, but these have $\pphi<0$.

\subsection{Geodesics at $r=0$}
We now look for geodesics that stay in the submanifold $r=0$. Since these geodesics are by definition trapped as they remain inside a bounded region and cannot escape to infinity, we will discuss whether or not there are geodesics that are restricted to some range of $\theta$ and if they can reach the evanescent ergosurface at $\theta=\pi/2$ (and $r=0$).

In a similar way to the geodesics in the submanifold  $\theta=\pi/2$, if $V_r(r_0)=0$ and $V_r'(r_0)=0$ at some $r_0=r(\lambda_0)$ then $\dot{r}(\lambda_0)=0=\ddot{r}(\lambda_0)$ and the geodesic stays at constant $r_0$. Substituting $r=0$ into $V_r$ in \eqref{eq:Vr} we see that $V_r|_{r=0}=0$ and $V_r'(0)=0$ as required if \begin{equation}
p_z=0,\qquad \ppsi=0,\qquad \frac{\Lambda}{p_t^2}=(\frac{\sqrt{\Qo \Qf}}{a}-\bphi)^2. \label{eq:rcond}
\end{equation}

We now investigate the sign of the angular potential \eqref{eq:Vth} to find the 'allowed' and 'forbidden' regions for the geodesics. Substituting the appropriate values \eqref{eq:rcond} into \eqref{eq:Vth} gives \begin{equation}
V_{\theta}=\frac{p_t^2}{\sin^2 \theta}\left[a^2\sin^4\theta-\left((\bphi-\frac{\sqrt{\Qo\Qf}}{a})^2+\Qo+\Qf+a^2\right)\sin^2\theta+\bphi^2\right] \label{eq:Vth0}.
\end{equation}

Define \begin{equation}
x=\sin^2\theta, \qquad x\in[0,1].
\end{equation} To find the regions where geodesics can exist we need to find how many roots of \begin{equation}
W_{\theta}(x)=a^2 x^2-\left((\bphi-\frac{\sqrt{\Qo\Qf}}{a})^2+\Qo+\Qf+a^2\right)x+\bphi^2
\label{eq:Wth}\end{equation} there are in the range $x\in [0,1]$. 

By examining the coefficients of \eqref{eq:Wth} we find that $W_{\theta}|_{x=0}\geq 0$ and that there are always two real roots $x_-,\,x_+$ that are both positive (and $x_-=x_+$ is only possible if $\Qo=\Qf$). We then have two possibilities:
\begin{enumerate}[i)]
	\item{$\bphi>\xi$.} This splits into two subcases, neither of which give geodesics in $r=0$:	\begin{itemize}
		\item{$\left(\bphi-\frac{\sqrt{\Qo\Qf}}{a}\right)^2>a^2-\Qo-\Qf$} \\
			Both roots $x_-,x_+>1$ and the effective potential is strictly positive for $x\in [0,1]$. This implies that there are no geodesics with this range of values of the impact parameter in the submanifold $r=0$.
		\item{$\left(\bphi-\frac{\sqrt{\Qo\Qf}}{a}\right)^2< a^2-\Qo-\Qf$} \\ This only happens if the background parameters of the microstate geometry satisfy \begin{equation*}
		\Qo+\Qf<\frac{\Qo\Qf}{a^2}+2\sqrt{\Qo\Qf}\left(1-\frac{\Qo+\Qf}{a^2}\right)^{\frac{1}{2}}.
		\end{equation*} If we divide this through by $a^2$ we have a constraint equation for the dimensionless $Q_i/a^2$ which we can plot and find it is in fact not possible to find values of $Q_i/a^2$ that satisfy this. 
	\end{itemize}
	\item{$\bphi\leq \xi $} \\ \hfill
	There is one root $x_-\in[0,1]$ and the other root $x_+\geq1$. The geodesics can exist in $\theta\in[\theta_-,\pi/2]$ and therefore are allowed in a region that includes the evanescent ergosurface.
\end{enumerate}

The zero energy null geodesics of \cite{Eperon:2016cdd} that are trapped exactly on the evanescent ergosurface can seen by taking $p_t\rightarrow 0$ in \eqref{eq:Vth0}. If $\pphi$ is negative this is the limit $\bphi\rightarrow -\infty$ and we have case (ii) with the root $x_-\rightarrow 1$ so the geodesics are trapped exactly at $\theta=\pi/2$. 

Geodesics in the submanifold $r=0$ can reach the evanescent ergosurface at $\theta=\pi/2$ when $\bphi\leq \xi $. We can compare this to geodesics in the equatorial plane, which reach $r=0$ when $C\leq 0$ in \eqref{eq:Vrpi2}: this happens when $\bphi\leq \xi$ so we in fact have the same range of $\bphi$ for which geodesics reach the evanescent ergosurface in either submanifold. However, the geodesics in the equatorial plane that can reach $r=0$ do not cross from $\theta=\pi/2$ to the submanifold $r=0$ (and $\theta$ arbitrary), which are in orthogonal surface. This can be seen either by noting that the geodesics in each submanifold have different separation constants and so the equivalent of the Carter constant is different for each family, or that there is an $S^3$ which shrinks to zero size at $r=0,\,\theta=\pi/2$ and in coordinates that are regular near the evanescent ergosurface it can be seen that this prevents the geodesics from crossing here. 

 \subsection{Geodesics at constant $r$}\label{sec:rconst}
We mentioned in section \ref{sec:trapping} that in the Kerr spacetime there are null geodesics that follow unstable circular orbits in the equatorial plane with radius $r_-$ or $r_+$. In Kerr, there are also spherical photon orbits that remain at constant $r$ but for which $\theta$ varies, and the radius $r$ of these orbits is in the range $r_-\leq r\leq r_+$ \cite{Teo2003}. We will look for an analogue of these spherical photon orbits in the 2-charge microstate geometries, and find that they do exist but that in contrast to Kerr there is in general no restriction on the radius of these orbits; however, if we set $p_z=0=\ppsi$ the constant-$r$ geodesics are restricted to the range $r_-\leq r \leq r_+$ where $r_-,\,r_+$ are the radii of the unstable trapped orbits in the equatorial plane.

A geodesic at constant $r=r_0$ must have $V_r(r_0)=0$ and $V_r'(r_0)=0$ so that, from \eqref{eq:req}, if $r(\lambda_0)=r_0,\,\dot{r}(\lambda_0)=0$ then $\dot{r}(\lambda)=0$ and $\ddot{r}(\lambda)=0$ and the geodesic remains at $r_0$ for all values of the affine parameter $\lambda$. The geodesic then takes values of $\theta$ for which $V_{\theta}\leq 0$; this must be the case for at least one value $\theta_0\in[0,\pi/2]$ for the geodesic to exist at all. We therefore find values $r_0$ such that $V_r(r_0)=0$, $V_r'(r_0)=0$ and $V_{\theta}(\theta)\leq 0$ for some values of the parameters $\bpsi,\,\bphi,\, b_z,\,\Lambda/p_t^2$ and some $\theta$.

Define $p_t^2 W_r=r^2 V_r$ so that $W_r$ is a cubic polynomial in $r^2$ and \begin{equation*}
\{W_r(r_0)=0,\,\dfrac{dW_r}{dr}(r_0)=0 \} \Rightarrow \{V_r(r_0)=0,\,\dfrac{dV_r}{dr}(r_0)=0\} \text{ or } r=0. \end{equation*}
For ease of notation, let \begin{equation}
\Lt=\Lambda/p_t^2, \qquad \mu=1-b_z^2\leq 1,\qquad \nu=a \bpsi+\sqrt{\Qo\Qf}b_z
\end{equation}

We will find values of $r$ for which it is possible to have geodesics at constant $r$ by using the equations \begin{equation}
W_r(r)=0,\qquad \dfrac{dW_r}{dr}=0 \label{eq:rconst}
\end{equation} to find the parameters $\Lt$ and $\bphi$ in terms of $r^2$, $\bpsi$ and $b_z$. Then substituting these into the requirement that $V_{\theta}\leq 0$ for some value of $\theta$ gives an inequality of the form $F(r^2,\bpsi,b_z,\theta)\geq 0$: given values of $\bpsi,\,b_z$ and $\theta$ we can use this to find the possible range of $r$.

Solving equations \eqref{eq:rconst} for $\Lt$ and $\bphi$ gives:
\begin{equation} \begin{split}
\Lt=& 2\mu r^2+\mu a^2+a^2\nu^2 \frac{1}{r^4}\\
\bphi=& \frac{\sqrt{\Qo\Qf}}{a}\pm (\frac{a}{r^2}+\frac{1}{a})\sqrt{\nu^2+\mu r^4}. \label{eq:lb}
\end{split}
\end{equation} Note that we must have $\nu^2+\mu r^4\geq 0$ so that $\bphi$ is real.
For the parameters to be such that $V_{\theta}\leq 0$ for some $\theta$, from \eqref{eq:Vth}, \begin{equation}
\Lt\geq \frac{\bphi^2}{1-u^2}+\frac{\bpsi^2}{u^2}-\mu(\Qo+\Qf)-\mu a^2  u^2
\end{equation} where $u=\cos \theta$, $0\leq u \leq 1$. Substituting in $\Lt(r^2,b_z,\bpsi)$ and $\bphi(r^2,b_z,\bpsi)$ from \eqref{eq:lb} gives the inequality \begin{equation} 
F(r^2,b_z,\bpsi,u)\geq 0 \label{eq:Fineq} \end{equation} where \begin{equation}
\begin{split}
F(r^2,b_z,\bpsi,u)=2\mu r^6+\mu (a^2+\Qo+\Qf)r^4+a^2\nu^2 -\frac{\nu^2}{1-u^2}\frac{(a^2+r^2)^2}{a^2} -\frac{1}{1-u^2}\frac{(a^2+r^2)^2}{a^2}\mu r^4\\ \mp \frac{1}{1-u^2}\frac{2\sqrt{\Qo\Qf}}{a^2}r^2(a^2+r^2)\sqrt{\nu^2+\mu r^4}-\frac{1}{1-u^2}\frac{\Qo\Qf}{a^2}r^4-\Big(\frac{(\nu+\Qo\Qf\sqrt{1-\mu})^2}{a^2u^2}-\mu a^2u^2 \Big)r^4 \label{eq:F}
\end{split}
\end{equation} which gives a constraint on the values of $r$ for which it is possible to have geodesics at constant $r$ for certain values of $\bpsi,\,b_z$ and $u$. Alternatively, \eqref{eq:Fineq} could be used to find the range of $\theta$ for a geodesic at constant $r$ given $r,\,\bpsi$ and $b_z$.

Instead of attempting to explicitly solve $F\geq 0$ to find the allowed radii of the constant-$r$ geodesics in terms of $b_z,\,\bpsi$ and $u$ we will simply explain why in general these geodesics can actually exist at any radius. In \cite{Eperon:2016cdd} it is found that there is are stably trapped geodesics with tangent $V$ through every point in the spacetime. As $V=\pd/\pd t+\pd /\pd z$ these geodesics remain at constant $r$ (and $\theta$) and hence must satisfy \eqref{eq:Fineq} with $\mu=0$ and $\bpsi,\,b_z$ as appropriate.

It is interesting to consider the case $\ppsi=0$ so that there are geodesics which lie entirely within the submanifold $\theta=\pi/2$. If we also set $p_z=0$ then we know the radii of the unstably geodesics in the equatorial plane from section \ref{sec:pi/2}. In this case we actually find something similar to the spherical photon orbits of Kerr \cite{Teo2003}: the geodesics at constant $r$ with $p_z=0,\,\ppsi=0$ can only exist in the range $r_-\leq r\leq r_+$ where $r_\pm$ are the radii of the unstable photon orbits in $\theta=\pi/2$ (it is possible that $r_-=0$). 

To see this, set $p_z=0=\ppsi$ in \eqref{eq:F} and multiply through by $(1-u^2)r^{-4}$, so the inequality $F\geq 0$ becomes \begin{equation}
G(u,\,r)=-a^2 u^4-(2r^2+\Qo+\Qf)u^2+\mathcal{Q}\geq 0
\end{equation} where we use the notation \begin{equation*}
\mathcal{Q}=2r^2+a^2+\Qo+\Qf-\frac{1}{a^2}\left(\sqrt{\Qo\Qf}\pm(a^2+r^2)\right)^2
\end{equation*} for comparison to the geodesics in Kerr \cite{Teo2003}. Observe that $G(1,r)<0$ and that the coefficient of $u^2$ is negative so that for geodesics to exist we must have $\mathcal{Q}\geq 0$ for $G(u,\,r)$ to be positive for some $u\in[0,\,1]$. This gives a range of allowed values for $r$ that we can calculate explicitly: $\mathcal{Q}$ is a quadratic polynomial in $r^2$ that is negative for $r^2\rightarrow \pm \infty$. We can therefore calculate the roots $r_{\pm}$ of $\mathcal{Q}$, and the allowed range of $r$ is between these roots. When $\mathcal{Q}=0$, $G(0,r_{\pm})=0$ and $u=0$ is a maximum of $G(u,r_{\pm})$: this implies that the geodesics are at $\theta=\pi/2$ and they are stable to perturbations in $\theta$. It turns out that the roots $r_{\pm}$ such that $\mathcal{Q}(r_{\pm})=0$ are precisely the radii of the unstable orbits in the equatorial plane, and so the geodesics at constant $r$ are only possible in the range $r_-\leq r\leq r_+$, with $r_{\pm}$ are given in \eqref{eq:radii}.

If we only set $p_z=0$ we again find that it is not possible to have constant-$r$ geodesics everywhere. Expanding $F$ for large $r$, one finds that there are no values for $\bpsi$ for which $F$ is positive; therefore for large $r$ there are no constant radius orbits.

\section{Null geodesics with momentum around the internal torus}\label{sec:timelike}
Similarly to Section \ref{sec:geo}, we will now investigate null geodesics in 10d that have non-zero momenta around the internal torus and which are timelike geodesics of the 6d metric $(\tilde{g}_6)_{\mu\nu}$ in \eqref{eq:gtilde} (\textit{not} of the Einstein frame metric). In particular we find that some of these geodesics are stably trapped, and these can then be used in the argument for instability in the Introduction that involves a massive particle.

The equations for these geodesics are given by the two coupled 1d equations of motion in \eqref{eq:theq} where $\mu$ is the mass of the particle (in 6d), $\sqrt{H_2/H_1}g_6^{\mu \nu}P_{\mu}P_{\nu}=-\mu^2$. The effective potentials are given in \eqref{eq:Vr} and \eqref{eq:Vth}.

\subsection{Geodesics in the equatorial plane}
In an analogous way to section \ref{sec:pi/2}, there are geodesics that stay in the equatorial plane $\theta=\pi/2$ if \begin{equation}
\ppsi=0,\qquad \Lambda=\pphi^2+(p_z^2-p_t^2)(\Qo+\Qf) \end{equation}
and for simplicity we will investigate the case $p_z=0$.

In this case the radial equation reduces to \begin{equation}
\dot{r}^2+\frac{1}{h^2f^2}U_r=0 \label{eq:radeq}
\end{equation} where \begin{equation}
\frac{1}{p_t^2}U_r=(m^2-1)r^4+(B+m^2(a^2+\Qf))r^2+(C+m^2a^2 \Qf)
\end{equation} and $B,\,C$ are given in \eqref{eq:Vrpi2} and $m=-\mu/p_t$.

We will briefly discuss whether or not it is possible to have stable trapping in terms of the binding energy. There are two cases: \begin{enumerate}[i)]
	\item Positive binding energy, $\mu>-p_t$. \\
	Geodesics cannot even exist near infinity and it is only possible to have bound orbits, so the only possibility is that the geodesics are stably trapped. If $\bphi\leq \xi-m^2a\sqrt{\Qf/4\Qo}$ the geodesics are trapped in a region which includes the evanescent ergosurface.
	\item Negative binding energy, $\mu\leq -p_t$. \\
	It is not obvious in this case that there are  stably trapped geodesics, but in fact there are geodesics that are stably trapped near the evanescent ergosurface; this can be seen by taking $\bphi\rightarrow \infty$. If we do this by leaving $p_t$ finite but taking $\pphi$ to be large and negative then the terms involving $\bphi$ are much larger than those which depend on $m$. Since the potential is then almost the same as for the null geodesics in 6d in section \ref{sec:pi/2}, we know there will be stable trapping for $\bphi$ sufficiently large and negative. We also expect there to be other ranges of the parameters that give rise to stable trapping.
\end{enumerate}

It is interesting to observe that, depending on the background parameters, for particles with positive binding energy it can be shown that it is possible to find geodesics in the equatorial plane that are stably trapped in a region that does not include the evanescent ergosurface at $r=0$. This is in contrast to the results of \cite{Tangherlini1963}, where it is shown that in a Schwarzschild spacetime in $n+1$ dimensions, if $n>3$ there are no stable bound orbits. Similarly, around a 5d Myers-Perry black hole there are no bound orbits outside of the event horizon \cite{Diemer:2014lba}, in contrast to the supersymmetric microstate geometries.

\section{3-charge microstate geometries}\label{sec:3charge}

In this section we will study null geodesics in the 3-charge microstate geometries of Refs. \cite{Giusto:2004id,Giusto:2004ip,Giusto:2004kj}. Similarly to the 2-charge microstate geometries in the previous section, these are supersymmetric solutions of type IIB supergravity compactified on $T^4$. The resulting 6d geometry asymptotically approaches the product of 5d Minkowski spacetime with a Kaluza-Klein circle of radius $R_z$. We will focus on the case where the 6d geometries are smooth with no conical or orbifold singularities. 

These solutions admit 4 Killing vector fields and a "hidden" symmetry (associated to a conformal Killing tensor field) which enables one to separate both the wave equation \cite{Giusto:2004ip} and the Hamilton-Jacobi equation for null geodesics into ODEs. 

The 3 charges of these solutions arise from $n_1$ D1-branes wrapped around the Kaluza-Klein $S^1$, $n_2$ D5-branes wrapped around $S^1\times T^4$ (the same as the 2-charge supersymmetric microstate geometry) but also with $n_p$ units of momentum around the $S^1$ where
\begin{equation}
n_p = n(n+1) n_1 n_2,\qquad n \in \mathbb{Z}.
\end{equation}
The solution is written in terms of dimensionful charges
\begin{equation}
\Qo=\frac{(2\pi)^4g\alpha'^3}{V}n_1 \qquad \Qf=g\alpha'n_2 \qquad Q_p=a^2n(n+1)=\frac{4G^{(5)}}{\pi R_z}n_p
\end{equation}
where $g$ is the string coupling constant, $V$ is the volume of the $T^4$, $G^{(5)}$ is the 5d Newton constant and the length scale $a$ is defined in \eqref{eq:a}.
The 10d string frame metric is:
\begin{equation} \begin{split}
ds^2=&-\frac{1}{h}(dt^2-dz^2)+\frac{Q_p}{hf}(dt-dz)^2+hf\Big(\frac{dr^2}{r^2+(\gto+\gtt)^2\eta}+d\theta^2\Big) \\ &+h\Big(r^2+\gto(\gto+\gtt)\eta-\frac{(\gto^2-\gtt^2)\eta Q_1Q_2\cos^2\theta}{h^2f^2}\Big)\cos^2\theta d\psi^2\\ &+h\Big(r^2+\gtt(\gto+\gtt)\eta+\frac{(\gto^2-\gtt^2)\eta Q_1Q_2\sin^2\theta}{h^2f^2}\Big)\sin^2\theta d\phi^2 \\ &+\frac{Q_p(\gto+\gtt)^2\eta^2}{hf}(\cos^2\theta d \psi+\sin^2\theta d\phi)^2 \\&-2\frac{\sqrt{Q_1Q_2}}{hf}\Big(\gto\cos^2\theta d\psi+\gtt \sin^2\theta d\phi\Big)(dt-dz)\\&-2\frac{(\gto+\gtt)\eta\sqrt{Q_1Q_2}}{hf}\Big(\cos^2\theta d\psi+\sin^2 \theta d\phi \Big)dz+\sqrt{\frac{H_1}{H_2}}\Sigma_{i=1}^{4}dx_i^2
\\=& ds^2_6+\sqrt{\frac{H_1}{H_2}}\Sigma_{i=1}^{4}dx_i^2 \end{split} \label{eq:metric} \end{equation}
where 
\begin{equation}
\eta=\frac{Q_1Q_2}{Q_1Q_2+Q_1Q_p+Q_2Q_p}, \label{eq:eta}
\end{equation}
\begin{equation}
\tilde{\gamma}_1=-an,\;\;\tilde{\gamma}_2 =a(n+1),\
\end{equation}
\begin{equation}
f=r^2+a^2\eta(-n\sin^2\theta+(n+1)\cos^2\theta),
\end{equation}
\begin{equation}
H_1=1+\frac{Q_1}{f},\;\;H_2=1+\frac{Q_2}{f}\; \text{ and } \; h=\sqrt{H_1H_2},
\end{equation}
where $\theta\in [0,\pi/2]$, $r>0$ and $0 \le \phi,\psi \le 2\pi$. 

The angular momenta of these geometries are 
\begin{equation}
J_{\psi}=- n n_1n_5, \qquad J_{\phi}=(n+1) n_1n_5.
\end{equation} 

The 3-charge solution reduces to the 2-charge supersymmetric microstate geometry in the previous section if we set $n=0$. 

The evanescent ergosurface is given by $f=0$, where the globally null Killing vector field \begin{equation} V=\dfrac{\pd }{\pd t}+\dfrac{\pd}{\pd z} \end{equation} is orthogonal to the Kaluza-Klein Killing vector field $Z=\pd/\pd z$. This gives the submanifold defined by \begin{equation*}
r^2=a^2\eta\left(n\sin^2\theta-(n+1)\cos^2\theta\right)
\end{equation*} which has topology $S^1\times S^3$ at constant $t$.

\subsection{Null geodesics}
The Hamilton-Jacobi equation for null geodesics \eqref{eq:HJ} in the supersymmetric 3-charge microstate geometries separates into coupled first order equations of motion. We can use the effective potentials to find regions where geodesics are either 'allowed' or 'forbidden' and from this we can see whether there are stable or unstable trapped null geodesics.

We will not restrict to any particular values of the conserved quantities or to any particular submanifold as we did for the 2-charge microstate geometry in section \ref{sec:2charge}. The results here are therefore more general, but we do not find exact bounds on the values of the impact parameters for which the different types of trapping occur.

\subsubsection{Equations of motion}
The behaviour of a geodesic depends on its conserved quantities. We will factor out the energy $-p_t$ because we can rescale the affine parameter along the null geodesics so only the ratios of the momenta have any physical meaning. We define the impact parameters
\begin{equation}
\bphi=-\frac{\pphi}{p_t}, \qquad \bpsi=-\frac{\ppsi}{p_t}, \qquad b_z=-\frac{p_z}{p_t}.
\end{equation}

Due to the extra symmetry associated to a conformal Killing tensor that allows one to separate the wave equation, the Hamilton-Jacobi equation for null geodesics also separates. If we substitute the ansatz \eqref{eq:S10} and the inverse of the metric \eqref{eq:metric} (which can be found in ref. \cite{Giusto:2004ip}) into the Hamilton-Jacobi equation \eqref{eq:HJ} we obtain first order equations for $R(r)$ and $\Theta(\theta)$. In the same way as for the 2-charge microstate geometry in section \ref{sec:2charge}, this allows us to find the coupled first order equations of motion for $r(\lambda)$ and $\theta(\lambda)$:
\begin{align} 
\dot{r}^2+\frac{1}{(hf)^2}U_r(r)=0 \label{eq:req3} \\ \dot{\theta}^2+\frac{1}{(hf)^2}U_{\theta}(\theta)=0
 \label{eq:theq3} \end{align}
where $\dot{}=\dfrac{d}{d\lambda}$, $\lambda$ is an affine parameter along the null geodesic. The effective potentials are
\begin{equation} \begin{split}
U_{\theta}=a^2\eta\Big(- n\sin^2\theta+& (n+1)\cos^2\theta\Big)(p_z^2-p_t^2+ \mu^2)+(\Qo+\Qf)(p_z^2-p_t^2)\\& -Q_p(p_t+p_z)^2+\frac{\ppsi^2}{\cos^2\theta}+\frac{\pphi^2}{\sin^2\theta}-\Lambda \label{eq:thpot}  \end{split} \end{equation}
\begin{equation} \begin{split}
U_r=\frac{p_t^2}{r^2}\Bigg[-\eta r^2\Big(\frac{\sqrt{\Qo\Qf}}{\eta}& -\frac{\Qo+\Qf}{\sqrt{\Qo\Qf}}a^2n(n+1)b_z-a(n+1)\bphi+an\bpsi\Big)^2+r^4(r^2+a^2\eta)(b_z^2-1)\\&+(r^2+a^2\eta)\eta\Big(-\sqrt{\Qo\Qf}b_z-a(n+1)\bpsi+an\bphi\Big)^2+\frac{\Lambda}{p_t^2}r^2(r^2+a^2\eta)\\ & +\frac{\mu^2}{p_t^2}r^2\big(r^4+(\Qf+a^2\eta)r^2+\Qf a^2\eta \big)\Bigg]\end{split} \label{eq:rpot} \end{equation}
where $\mu^2=\delta^{ij}q_iq_j$, $i,\,j=1\dots 4$, $q_i$ is the conserved momenta around the internal torus and $\Lambda$ is a constant arising from the separation of variables. Note that the coupling factor $(hf)^{-2}$ is strictly positive so the 'allowed' regions for geodesics only depend on the signs of the effective potentials.

\subsubsection{Trapped geodesics}
We will describe the possibilities for trapping when $\mu=0$, since these correspond to null geodesics in 6d.

Geodesics can only exist in regions where both of the effective potentials are negative. Therefore there must be some value $\theta\in[0,\pi/2]$ such that $U_{\theta}$ is negative; this implies that the separation constant $\Lambda$ must satisfy \begin{equation}
\Lambda\geq a^2\eta\Big(-n\sin^2\theta+(n+1)\cos^2\theta\Big)(p_z^2-p_t^2) +\frac{\ppsi^2}{\cos^2\theta_0}+\frac{\pphi^2}{\sin^2\theta_0} \label{eq:thcond} 
\end{equation}
for some $\theta_0\in [0,\pi/2]$ so that the geodesic can exist at least at $\theta_0$. 

If we let $x=\cos\theta$ we can write $x^2(1-x^2)U_{\theta}(x)$ as a cubic polynomial in $x^2$ which is positive at $x=0$ and $x=1$. Therefore there can be at most one region in the range $0\leq x \leq 1$ where geodesics are allowed; it is not possible to have disjoint regions for the geodesics for the same values of the impact parameters.

Assuming we choose $\Lambda$ such that \eqref{eq:thcond} holds, the problem is then to find the regions in $r$ where geodesics can propagate, i.e. where $U_r$ is negative. To do this we consider the terms in square brackets in $U_r$ in \eqref{eq:rpot}, which give a cubic polynomial in $r^2$ and so can have up to 3 positive roots. We write this expression as \begin{equation}
W_r=(b_z^2-1)r^6+\alpha r^4+\beta r^2+\gamma
\end{equation}
for some $\alpha,\,\beta,\,\gamma$ that are given by the coefficients in \eqref{eq:rpot} and depend on $\bphi,\,\bpsi,\,b_z$ and $\Lambda/p_t^2$, although it is useful to note that for any values of the parameters $\gamma\geq 0$. As $W_r$ depends on these parameters, whether or not there are trapped geodesics also depends on $\bphi,\,\bpsi,\, b_z$ and $\Lambda/p_t^2$.

\begin{figure}[h]
\centering
\includegraphics[width=0.7\linewidth]{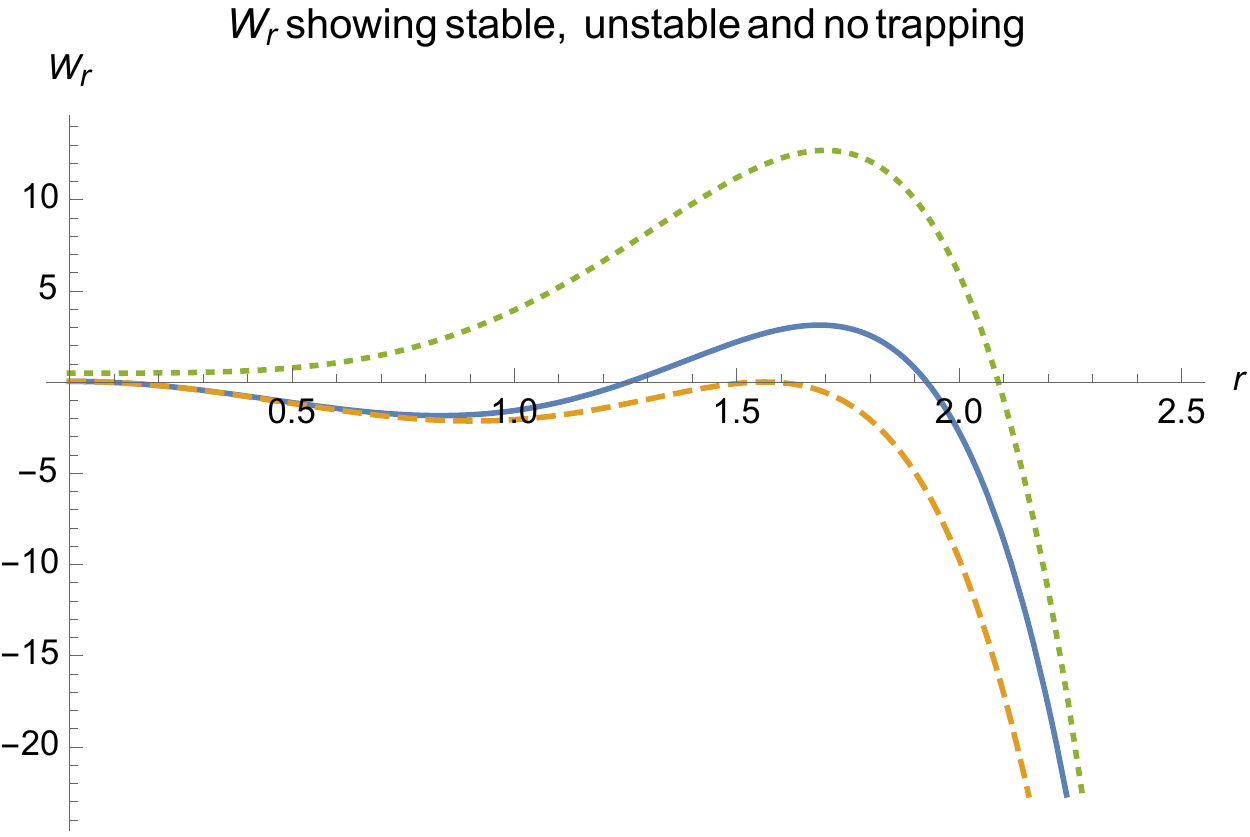}
\caption{Plot of $W_{r}$ showing that for different values of the impact parameters there is stable trapping (solid blue line) , unstable trapping (dashed orange line) or no trapping (dotted green line).}
\label{fig:trapping}
\end{figure}

If $p_z^2-p_t^2<0$, which in particular includes geodesics with zero Kaluza-Klein momentum, $W_r$ can have up to 3 positive real roots: the potential is positive at $r=0$ because $\gamma\geq 0$ and  $p_z^2-p_t^2<0$ implies that $W_r(r)\rightarrow-\infty$ as $r\rightarrow\infty$. We therefore have the following possibilities for geodesics:
\begin{enumerate}[i)]
	\item{$W_r$ has 3 positive real roots, $\{r_-,\,r_0,\,r_+\}$.} \\ \hfill
	If all the roots are distinct, $r_-<r_0<r_+$, then for the values of $\bphi,\,\bpsi,\,p_z/p_t$ and $\Lambda/p_t^2$ that allow for such roots there are stably trapped geodesics in the region $r_-\leq r\leq r_0$ as well as geodesics only allowed in $r\geq r_+$ that can escape to infinity. 
	
	If $r_0=r_+$ there is unstable trapping at $r_0$. This might occur for several different values of $\bphi,\,\bpsi,\,\frac{p_z}{p_t}$ and $\Lambda/p_t^2$; as $r_0$ depends on these parameters this unstable trapping could occur at various values of $r$.
	
	If $r_-=r_0<r_+$ there is stable trapping with the geodesics localised exactly at $r_0$.
	\item{$W_r$ has 1 real root, $r_0$.} \\ \hfill
	For these values of $\bphi,\,\bpsi,\,b_z$ and $\Lambda/p_t^2$ there are only geodesics in the region $r\geq r_0$ which can escape out to infinity and therefore no trapping.
\end{enumerate}
Figure \ref{fig:trapping} plots $W_r$ in these cases with appropriate values of $\bphi,\,\bpsi,\,p_z/p_t$ and $\Lambda/p_t^2$. On this plot the 'allowed' and 'forbidden' regions for these geodesics are clear and it is easy to see whether or not the geodesics are trapped.  

If $p_z^2-p_t^2>0$ there can either be two distinct positive roots, one double root or $W_r$ is always positive. If there are two roots $r_-<r_+$ the geodesics are stably trapped in the region $r_-\leq r \leq r_+$. If $r_-=r_+$ the geodesics are stably trapped at exactly $r_-$. The only other possibility is that there are no real roots so $W_r$ is always strictly positive and no geodesics can exist.

Finally, if $p_z^2=p_t^2$ there are stably trapped geodesics if $\bphi,\,\bpsi,\,b_z$ and $\Lambda/p_t^2$ are such that $\alpha>0$, $\beta^2\geq 4\alpha\gamma$ and $\beta\leq 0$. If instead $\bphi,\,\bpsi,\,b_z$and $\Lambda/p_t^2$ give $\alpha\leq 0$, the geodesics only exist in the region $r\geq r_0$ for some $r_0$; if neither of these is the case then there are  no geodesics with those values of the parameters.

Note that for all of these cases to happen it must be possible to find roots of $W_r$ for some values of the impact parameters subject to the restriction on $\Lambda$ in \eqref{eq:thcond} which ensures there is an 'allowed' value of $\theta$ for the geodesic. Even in the cases where there is stable trapping there is never more than one region in which the geodesics can be trapped for a given set of parameters.

\subsubsection{Trapping and the evanescent ergosurface}
The same argument from \cite{Eperon:2016cdd}, given in section \ref{sec:2charge} for the 2-charge case, for the stable trapping of the zero-energy null geodesics also applies to these 3-charge microstate geometries. Therefore there are zero-energy null geodesics with tangent vector $V$ that are stably trapped on the evanescent ergosurface. 

This can be seen using the equations of motion and the effective potentials by setting $p_t=0$ and $p_z=0$ in \eqref{eq:rpot} and \eqref{eq:thpot}. In this case there is a minimum of $U_{\theta}$ at $\theta_0$ where $\tan^2 \theta_0=\pphi/\ppsi$, and by an appropriate choice of $\Lambda $ these geodesics are localised exactly at $\theta_0$. Substituting this into the radial potential, we find that there is a minimum at $r_0^2=a^2\eta|(n+1)\cos^2\theta-n \sin^2\theta|$ and $U_r'(r_0)=0$ so the geodesics are stably trapped exactly on the evanescent ergosurface. 

In general, the trapped region does not always include the evanescent ergosurface. Indeed, it was shown in \cite{Eperon:2016cdd} that there are stably trapped geodesics with tangent $V$ through every point in the spacetime. If these geodesics have non-zero energy they are not localised on the evanescent ergosurface, and these have $P_z\neq 0$ in general.

\subsection{Penrose process}
The Penrose process \cite{Penrose} is a method of extracting energy from a Kerr black hole. It is possible because there is an ergoregion, where the Killing vector field $T$, which is timelike at infinity, becomes spacelike outside of the event horizon. This means that the energy of a physical particle with momentum $P$ ($P$ is future-directed and causal), which is given by $E=-T\cdot P$, can become negative in the ergoregion. In the Kerr spacetime, it is possible to set up a situation in which a particle with positive energy in the ergoregion decays into two other particles that follow geodesics, one with negative energy that falls into the black hole and one with energy greater than that of the initial particle that escapes back out to infinity, thus extracting energy from the black hole.

There is also an ergoregion in the 3-charge microstates geometries, so it interesting to ask whether a similar process can happen here. We have only considered geodesics that are null in the full 10 dimensions; within this class of geodesics it is not possible to replicate the Penrose process exactly to find one particle that decays into two other particles since momentum conservation (in 10d) would require writing one null vector as a sum of two non-parallel null vectors, which is not possible.

Instead we will look for two particles sent in from infinity with positive energy interacting within the ergoregion to produce two other particles, one with negative energy that stays trapped in some region and one that escapes to infinity with energy greater than the sum of the energies of the two initial particles. To set up the initial conditions for this to happen, the particles that interact should be able to fall in from infinity, and all the particles should follow geodesics.

Note that the ergoregion is defined to be the region where $T^2>0$, which is the region where $f<Q_p$ (note that the evanescent ergosurface lies within the ergoregion). To simplify matters, we can restrict attention to geodesics that remain in the equatorial plane $\theta=\pi/2$ by setting \begin{equation}
\ppsi=0,\qquad \Lambda=-a^2\eta n(p_z^2-p_t^2+\mu^2)+(\Qo+\Qf)(p_z^2-p_t^2)-Q_p(p_t+p_z)^2+\pphi^2
\end{equation} in equation \eqref{eq:thpot}. If we substitute this into \eqref{eq:rpot} we find the radial effective potential for geodesics in the equatorial plane.

For our Penrose-like process, let the two particles with positive energy that interact and could come in from infinity have momenta $P$ and $Q$. Let the particle with negative energy that becomes trapped have momentum $R$ and the particle that escapes out to infinity with greater energy than the initial energy have momentum $S$. By momentum conservation, we must have \begin{equation}
P+Q=R+S. \label{eq:momcon3}
\end{equation}

In this example of energy extraction, we will set $a=1,\,\Qo=1=\Qf$ and $n=1$. Let $\mu_P=\sum_{i=1}^4q_i^2$ be the sum of the components of the momentum around the internal torus and $E_P$ (for example) be the energy of the particle with momentum $P$, so $E_P=-T\cdot P=-P_t$ and so by \eqref{eq:momcon3}, $E_P+E_Q=E_R+E_S$.

All of $P,\,Q,\,R$ and $S$ must be future-directed and causal. Recall that the Killing vector field $V=\pd/\pd t+\pd/ \pd z$ is future-directed and globally null, so the momentum of a physical particle must satisfy $V\cdot P\leq 0$, i.e. $P_t+P_z\leq 0$ to ensure it is future-directed. 

Since all the geodesics are in the equatorial plane, the $\theta,\,\psi-$components of the momenta are zero. From the equation for geodesics \eqref{eq:req3} we can find the radial components of the momenta in terms of the conserved $t,\,\phi,z-$components: \begin{equation}
(P^r)^2=\frac{-U_r(r)}{h^2f^2}. \label{eq:pr}
\end{equation}

In the equatorial plane, the ergoregion is given by $r^2<6/5$. We will assume that the process happens well inside the ergoregion, at $r=0.2$. We are free to pick most of the components of the momenta then use momentum conservation and the equation for the radial component to find the other. We do this by specifying the conserved components of $P$ (particle follows a geodesic that comes in from infinity), $R$, the particle with negative energy that becomes trapped and all but one of the conserved components of $S$ (the other is calculated from the condition arising from \eqref{eq:momcon3} and \eqref{eq:pr}). Equation \eqref{eq:momcon3} then gives $Q$. 

\begin{figure}
\centering
\includegraphics[width=0.7\linewidth]{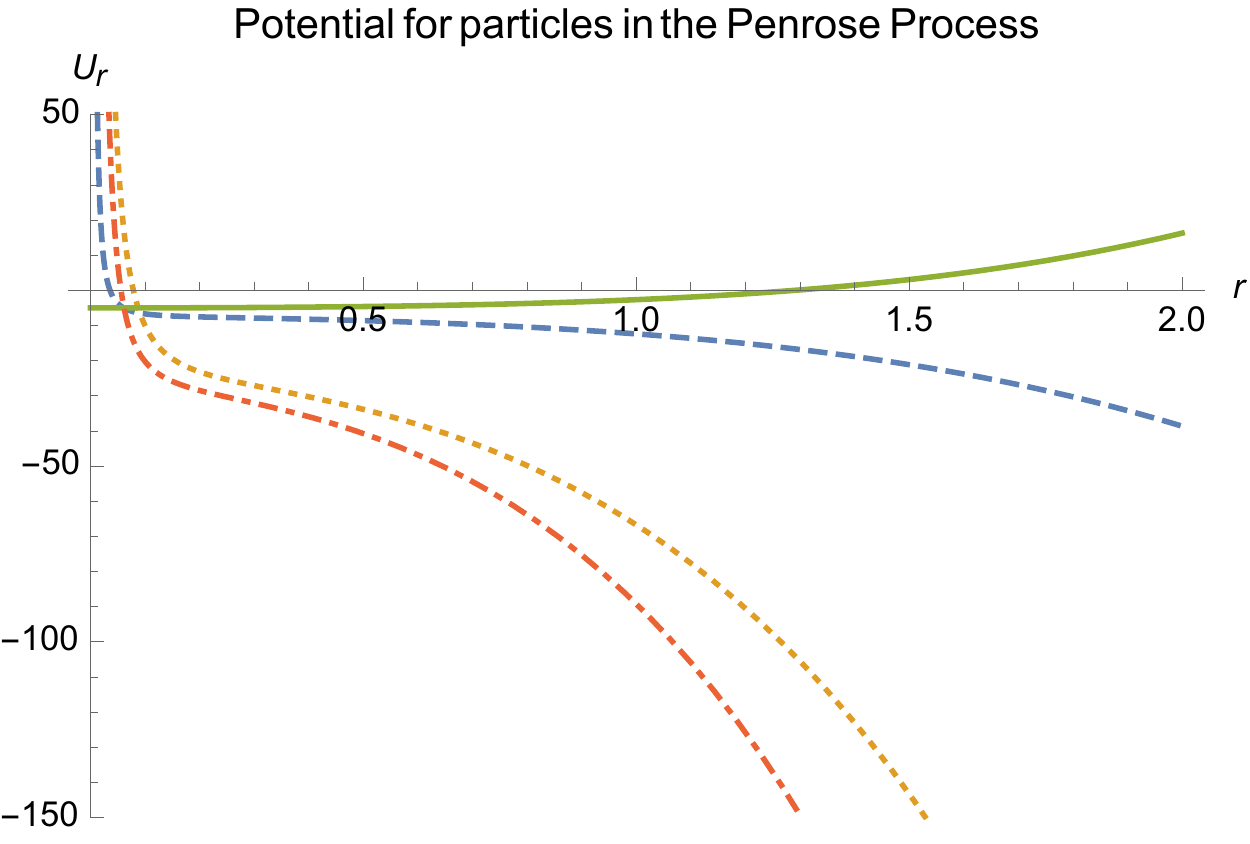}
\caption{The potentials for the particles above: the green solid line is for the particle $R$ that becomes stably trapped with negative energy, while the blue dashed, orange dotted and red dot-dashed are for $P,\,Q$ and $S$ respectively.}
\label{fig:penrose3}
\end{figure}

We can do this with particles that are null in 6d, so with $\mu=0$. Values of the momenta of the particles that allow such a process are as follows:
\begin{equation} \begin{split}
P_t=-1,\qquad P_{\phi}=-\frac{1}{2},\qquad P_z=0 \qquad \Rightarrow hf|_{r=0.2}P^r =2.75\dots \\
 Q_t=-3 ,\qquad Q_{\phi}=2.02\dots,\qquad Q_z=0 \qquad \Rightarrow hf|_{r=0.2}Q^r=4.82\dots \\
R_t=0.1,\qquad R_{\phi}=-1,\qquad R_z=-1 \qquad \Rightarrow hf|_{r=0.2}R^r=2.23\dots \\
S_t=-4.1,\qquad S_{\phi}=2.52\dots ,\qquad S_z=1\qquad \Rightarrow hf|_{r=0.2} S^r=5.34\dots.
\end{split}
\end{equation}
The potentials $U_r(r)$ corresponding to each of these particles are shown in Figure \ref{fig:penrose3}.

Note that the potential for $R$ shows that the particle is stably trapped in some region since it is positive at infinity, and that $E_R=-0.1<0$ so this particle is indeed trapped with negative energy. The particle with momentum $S$ has energy $E_S>E_P+E_Q$ as required and it can be seen that it can escape back out to infinity. 

It is interesting to observe that, despite the ergoregion and the existence of timelike geodesics with negative energy within it, there is no Friedman instability \cite{friedman1978}. One might expect that, due to the ergoregion, there could be solutions of the Klein-Gordon equation which are not uniformly bounded and indeed even grow. However, this should be prevented for the same reason that all solutions of the massless wave equation are bounded in these geometries despite the presence of the ergoregion \cite{Keir:2016azt}: there is a globally null Killing vector field $V$ that provides us with a conserved, but degenerate, energy that we can use to bound the non-degenerate energy.

\section{Implications for quasinormal modes}
Quasinormal modes are mode solutions of the wave equation $\Box \Phi=0$ of the form \begin{equation}
\Phi(t,x^i)=e^{-i \omega t}\Psi(x^i),
\end{equation} where $\omega=\omega_R+i\omega_I$ has both real and imaginary parts and is known as the quasinormal frequency. Quasinormal modes must be outgoing at infinity. The inner boundary condition depends on whether or not there is a horizon. If there is, the solution must be ingoing at the horizon; if not, as is the case for the microstate geometries, we only require that the solution is regular everywhere. These boundary conditions give rise to a discrete set of quasinormal frequencies.

If there is enough symmetry in the spacetime and the wave equation separates, we can write a solution of the wave equation as 
\begin{equation}
\Phi(t, r, \theta, \phi, \psi, z)=e^{-i \omega t+im_{\psi}\psi+im_{\phi}\phi+i\lambda z}e^{i q_jz^j} \Phi_r(r)\Phi_{\theta}(\theta). \label{eq:qnm}
\end{equation}

Using the geometric optics approximation, we expect to be able to find rapidly varying solutions of the massless wave equation that are localised near null geodesics. This corresponds to taking the limit $|\mphi|+|\mpsi|\rightarrow \infty$; in this case $\Phi_{\theta}$ are approximately the angular harmonics labelled by an integer $\ell\geq |\mphi|+|\mpsi|$. We will assume in the following that $|\mphi|,|\mpsi|=O(\ell)$ and take the limit $\ell\rightarrow\infty$. 

We first discuss examples where the quasinormal modes in this limit have been calculated explicitly. We can then describe what we expect for quasinormal modes as a consequence of the geodesics we have found.

Around a Kerr black hole there are unstable circular photon orbits for which the geodesics have non-zero energy. In the limit $\ell\rightarrow \infty$ there are quasinormal mode solutions localised near these null geodesics with $\omega_R=O(\ell)$ and $\omega_I=O(1)$, $\omega_I<0$ \cite{Yang:2012he}.

In ultracompact neutron stars, which are fluid objects with a photon sphere but no horizon, and in Kerr-AdS there are stably trapped null geodesics with non-zero energy. For Kerr-Ads this leads to quasinormal modes with $\omega_R=O(\ell)$ and $\omega_I=O(e^{-\gamma \ell})$ for some positive constant $\gamma$ and $\omega_I<0$ \cite{Gannot2014,Dias:2012tq}. This stable trapping is the reason that the rate of decay of solutions to the wave equation is very slow in both Kerr-AdS and ultracompact neutron stars, which was proved in \cite{Holzegel:2013kna} and \cite{Keir:2014oka} respectively.

In the supersymmetric 2 and 3-charge microstate geometries that we discussed in the previous section, there are quasinormal modes localised near the zero-energy null geodesics that are stably trapped on the evanescent ergosurface that have $\omega_R=O(1)$ (this is because the geodesics have zero energy) and $\omega_I=O(e^{-2\ell \log \ell})$, $\omega_I<0$ \cite{Eperon:2016cdd}. There are also quasinormal modes localised near the geodesics with tangent $V$ that are not on the evanescent ergosurface and thus have non-zero momentum around the Kaluza-Klein direction. These have $(\omega_R-\lambda)=O(1)$ in \eqref{eq:qnm} and $\omega_I=O(e^{-\ell \log \ell})$, $\omega_I<0$. It was shown in \cite{Eperon:2016cdd} and proved rigorously in \cite{Keir:2016azt} that this leads to even slower decay for solutions of the wave equation than in the cases where the stably trapped geodesics have non-zero energy.

We now summarise the other geodesics we have found in the supersymmetric microstate geometries and the implications for quasinormal modes. In all the following we assume $\omega_I<0$ since the modes eventually disperse out to infinity, and that $\alpha_i>0$ are some positive constants. Some of the possible cases are as follows:
\begin{itemize}
	\item From section \ref{sec:pi/2} we know there are geodesics with zero momentum around the internal torus ($q_i=0$ and also $\lambda=0$ in \eqref{eq:qnm}) that have non-zero energy and can either be stably or unstably trapped. From the discussion above, we expect there to be quasinormal modes localised near the stably trapped null geodesics with $\omega_R=O(\ell),\, \omega_I=O(e^{-\alpha_1 \ell})$. In case 1 of section \ref{sec:pi/2} the stably trapped geodesics can be either pro- or retrograde; this implies that $\mphi/\omega$ can be either positive or negative. However, in case 2 all the stably trapped null geodesics are retrograde, so we expect the quasinormal modes to have $\mphi/\omega<0$.
	
	We also expect there to be quasinormal modes with $\omega_R=O(\ell),\, \omega_I=O(1)$ localised near the unstably trapped null geodesics. This also applies for the geodesics at constant $r$ in section \ref{sec:rconst}, but we expect these quasinormal modes to be localised near a single value of $r$ as opposed to some finite region. In case 1 of section \ref{sec:pi/2} the unstably trapped geodesics are retrograde so for these quasinormal modes we expect $\mphi/\omega<0$, whilst in case 2 these geodesics can be either pro-or retrograde so $\mphi/\omega$ can have either sign. 
	\item If the momentum around the internal torus is non-zero, we showed in section \ref{sec:geo} that we can always find stably trapped null geodesics in 10d. In terms of quasinormal modes, this means that there are solutions with $q_i\neq 0$ in \eqref{eq:qnm} that are localised in some finite region.  In section \ref{sec:timelike} we found stable and unstable trapped null geodesics (in 10d); we might expect that there are quasinormal modes localised near these with $\omega_R=O(\ell)$ and $\omega_I=O(e^{-\alpha_2 \ell})$ or $\omega_I=O(1)$ respectively.
\end{itemize}
We expect analogous consequences due to the stably and unstably trapped null geodesics found in section \ref{sec:3charge} for quasinormal modes in the 3-charge supersymmetric microstate geometries.

\section*{Acknowledgements}
I would like to thank my supervisor Harvey Reall for suggesting the project and for numerous helpful discussions. I am also grateful to Jorge Santos for reading a preliminary version of this paper.

\bibliographystyle{h-physrev5}
\bibliography{Geodesicsfinal}

\end{document}